\documentclass[epj]{svjour}

\usepackage{xspace,pstricks,graphics}
\usepackage{cases}

\newcommand{\ie}{{\it i.e.}\xspace}
\newcommand{\eg}{{\it e.g.}\xspace}

\newcommand{\ave}[1]{\left\langle #1 \right\rangle}

\newcommand{\elabel}[1]{\label{eq:#1}}
\newcommand{\eref}[1]{(\ref{eq:#1})}
\newcommand{\Eref}[1]{Eq.~(\ref{eq:#1})}
\newcommand{\Ereffirst}[1]{Equation~(\ref{eq:#1})}

\newcommand{\slabel}[1]{\label{sec:#1}}

\newcommand{\Sref}[1]{Section~\ref{sec:#1}}

\newcommand{\flabel}[1]{\label{fig:#1}}
\newcommand{\fref}[1]{Fig.~\ref{fig:#1}}
\newcommand{\Fref}[1]{Figure~\ref{fig:#1}}

\newcommand{\text}[1]{{\textrm{#1}}}

\newcommand{\GC}{\mathcal{G}}

\newcommand{\GCtilde}{\tilde{\GC}}

\newcommand{\OC}{\mathcal{O}}
\newcommand{\tautilde}{\tilde{\tau}}
\newcommand{\atilde}{\tilde{a}}
\newcommand{\upCut}{s_{\text{\scriptsize c}}}
\newcommand{\loCut}{s_{\text{\scriptsize 0}}}

\newcommand{\result}[1]{\emph{#1}}

\newcommand{\dint}[1]{\!\!\!\mathrm{d}#1~}

\newcommand{\perc}{{\text{\tiny perc}}}

\begin{document}
  
\title{On the scaling of probability density functions with apparent
power-law exponents less than unity}
\titlerunning{Power-law less than unity}

\author{Kim Christensen\inst{1,2}
       \and 
       Nadia Farid\inst{2} \and 
       Gunnar Pruessner\inst{2,3}\thanks{Present address:
           Department of Mathematics, Imperial College London,
           180 Queen's Gate, London SW7 2AZ, United Kingdom}
       \and
       Matthew Stapleton\inst{2}}
\institute{Institute for Mathematical Sciences, Imperial College London, 
	   53 Prince's Gate, London SW7 2PG, United Kingdom
           \and
           Blackett Laboratory, Imperial College London,
           Prince Consort Road, London SW7 2AZ, United Kingdom 
	   \and
           Mathematics Institute, University of Warwick,
           Gibbet Hill Road, Coventry CV4 7AL, United Kingdom
	   }
\authorrunning{K. Christensen {\it et al.}}
\date{\today}

\abstract{
   We derive general properties of the finite-size scaling of
   probability density functions and show that when the apparent
   exponent $\tautilde$ of a probability density is less than $1$, the
   associated finite-size scaling ansatz has a scaling exponent $\tau$
   equal to $1$, provided that the fraction of events in the universal
   scaling part of the probability density function is non-vanishing in
   the thermodynamic limit.
   We find the general result that $\tau \ge 1$ and
   $\tau \ge \tautilde$.
   Moreover, we show that if the scaling function $\GC(x)$ approaches a
   non-zero constant for small arguments, $\lim_{x \to 0} \GC(x) > 0$, then $\tau = \tautilde$.
   However, if the scaling function vanishes for small arguments,
   $\lim_{x \to 0} \GC(x) = 0$, then $\tau = 1$, again assuming a
   non-vanishing fraction of universal events.
   Finally, we apply the formalism developed to examples from the literature,
   including some where misunderstandings of the theory of scaling have led to erroneous conclusions.
}
  
\maketitle
\section{Introduction}  
Power-law probability densities are pervasive in the literature on critical and scale-invariant
systems \cite{Stanley:1971,StaufferAharony:1994,Jensen:1998}.
The supposedly sole scaling parameter in a finite system
at the critical point is the {\em upper cutoff} $\upCut$, diverging in the
thermodynamic limit, $L \to \infty$, and the
probability density function (PDF) $P(s;\upCut)$
for an {\em observable} (an event size or the order parameter) 
$s \in [0,\infty[$ obeys simple \emph{finite-size scaling} (FSS), that is,
\begin{equation}
P(s;\upCut) =a s^{-\tau} \GC \left(s/\upCut\right) \quad \text{for $s, \upCut \gg \loCut$},
\elabel{FSS}
\end{equation}
where $\loCut$ is a constant {\em lower cutoff} of the PDF.
\Ereffirst{FSS} is valid only in the region $s \gg \loCut$
where the corrections to scaling
are negligible \cite{Wegner:1972}. The dimensionful parameter $a$ is a
so-called non-universal
{\em metric factor} \cite{PrivmanHohenbergAharony:1991}, $\tau$ is a universal 
(critical) {\em scaling exponent}, 
$\GC$ is the universal {\em scaling function} that decays sufficiently fast for $s \gg \upCut$
to ensure that, as one would expect, all moments of the PDF are finite for finite system size, $L < \infty$.
The upper cutoff $\upCut$ is the \emph{characteristic size} of $s$ in a finite system.
Usually $\upCut=b L^D$ to leading order, where $b$ is another non-universal
{\em metric factor} and $D$ is the universal {\em spatial dimensionality} of the observable $s$.
Again, in general, there are sub-leading orders, that is, corrections to scaling of the form
$\upCut = bL^D\left(1 + b_1L^{-\omega_1} + b_2L^{-\omega_2} + \cdots\right)$
with $0 < \omega_1 < \omega_2 < \cdots$, which can be safely ignored in the analysis of the asymptotes.
The scaling exponent $\tau$ is \emph{uniquely defined} by \Eref{FSS} and
is associated with the power-law decay of
the distinctive onset of the rapid decay $P(\upCut; \upCut)=a \GC(1) {\upCut}^{-\tau}$ as a
function of $\upCut$.
If data are consistent with the FSS ansatz \eref{FSS}, we can perform a {\em data collapse}:
By plotting the transformed PDF $s^{\tau} P(s;\upCut)$ vs. the 
rescaled observable $s/\upCut$, all data for $s,\upCut \gg \loCut$ collapse onto the same curve representing 
the scaling function $\GC$, see \Fref{Fig1}.
However, due to the presence of the metric factor $a$, the scaling function $\GC$ is
determined by \Eref{FSS} only up to a prefactor.
This ambiguity can be resolved, for example by fixing
$\GC(1) = 1$ \cite{LubeckWillmann:2005}.

The upper and lower cutoffs 
define a \emph{scaling region} $\loCut \ll s \ll \upCut$ where the PDF shows
essentially power-law behaviour, rather informally
\begin{equation}
P(s;\upCut) \propto s^{-\tautilde} \quad \text{for $\loCut \ll s \ll \upCut$},
\elabel{power-law}
\end{equation}
which means that $s^{-\tautilde}$ is the leading order of $P(s;\upCut)$ in the 
region where $s \gg \loCut$ coexists with $s \ll \upCut$.
The \emph{apparent exponent} $\tautilde$ is the slope of a straight-line fit
to the PDF data in the
scaling region $\loCut\ll s\ll\upCut$ 
when plotting $\log P(s;\upCut)$ vs. $\log s$, see \Fref{Fig1}(a).

Let us reflect upon 
the FSS ansatz \eref{FSS} and the informal \Eref{power-law}.
\Ereffirst{power-law} is informal for various reasons. First, the
``$\propto$'' suggests a
proportionality without specifying what parameters a possible prefactor
would be allowed to depend on. 
Second, the upper cutoff appears in the condition $s \ll \upCut$, but $\upCut$ does not
enter \Eref{power-law} explicitly as it does in \Eref{FSS} through its role as a characteristic
scale in the dimensionless argument of the scaling function $\GC$. 
Finally, and most importantly, the scaling function $\GC$ is not present.
Hence, to derive \Eref{power-law} from \Eref{FSS}, one has to make
assumptions about the behaviour
of $\GC$ in the scaling region $\loCut \ll s \ll \upCut$. If the
scaling function is ``almost constant'', then $\tau = \tautilde$.
However, if the scaling function is a power law itself, then $\tau \neq \tautilde$.
To illustrate that, let us define a {\em cutoff function} $\GCtilde$,
that, by definition, converges to a non-vanishing value for small arguments and decays rapidly
for large arguments, and assume that the scaling function is of the form $\GC(x) = x^{\alpha} \GCtilde(x)$.
Since the cutoff function $\GCtilde$ is constant and non-zero 
in the scaling region, 
the PDF behaves like $a s^{-\tau} \left(s/\upCut\right)^\alpha \GCtilde(0)$ for $\loCut \ll s \ll \upCut$, so that
the scaling exponent $\tau$ in \Eref{FSS}
is related to the apparent
exponent $\tautilde$ in \Eref{power-law} via $\tau -\alpha= \tautilde$.

One of the key problems found in the literature is the identification of the scaling exponent $\tau$
with the apparent exponent $\tautilde$; however, they differ if the scaling function has a
power-law dependence on its argument.
First, we illustrate the consequences of such a scaling
function further, concluding that if the apparent exponent $\tautilde$ in \Eref{power-law} is less than unity,
then the scaling exponent $\tau$
in \Eref{FSS} is exactly $1$. Next, this result is derived in general
by putting the formalism, Eqs.~\eref{FSS} and \eref{power-law}, on
a sound mathematical basis, generating further general properties in the process such as
$\tau \geq 1$, $\tau \ge \tautilde$ and $\tautilde \ge 1 \Rightarrow  \tau = \tautilde$.
Finally, we review examples from the
literature that illustrate the formalism developed, including some that
suffer from a confusion of $\tau$ and $\tautilde$.

\subsection{Example}
To demonstrate the first result explicitly, we assume that the PDF
for the observable $s \in [\loCut,\infty[$ has the form
\begin{equation}
P(s;\upCut) = \atilde s^{-\tautilde}  \GCtilde\left(s/\upCut\right)
\quad\text{for $s \geq \loCut = 1$},
\elabel{measured:Ps}
\end{equation}
see \fref{Fig1}(a).
Motivated by a previous study \cite{FaridChristensen:2006}, 
in this example, the cutoff function is assumed to be
$\GCtilde(x)=(1-x)^{\tautilde} \theta(1-x)$ with $\theta$ being
the Heaviside step function. (A simpler, but more artificial example would
be to consider a cutoff function identical to a pure Heaviside step function.)
Along the lines of the considerations
above, this cutoff function converges to unity for small arguments and
vanishes for all $x>1$.
The apparent exponent, the slope of $P(s;\upCut)$ in the scaling
region as shown in \fref{Fig1}(a), therefore is the exponent
$\tautilde$ in \Eref{measured:Ps}.
A na{\"i}ve comparison of \Eref{FSS} and \Eref{measured:Ps} suggests
that the scaling exponent $\tau$ is the slope of a straight-line
fit to $P(s;\upCut)$ vs. $s$ in a double logarithmic plot,
but this is incorrect if $\tautilde < 1$, as we will see in the
following.

\begin{figure}
\includegraphics*{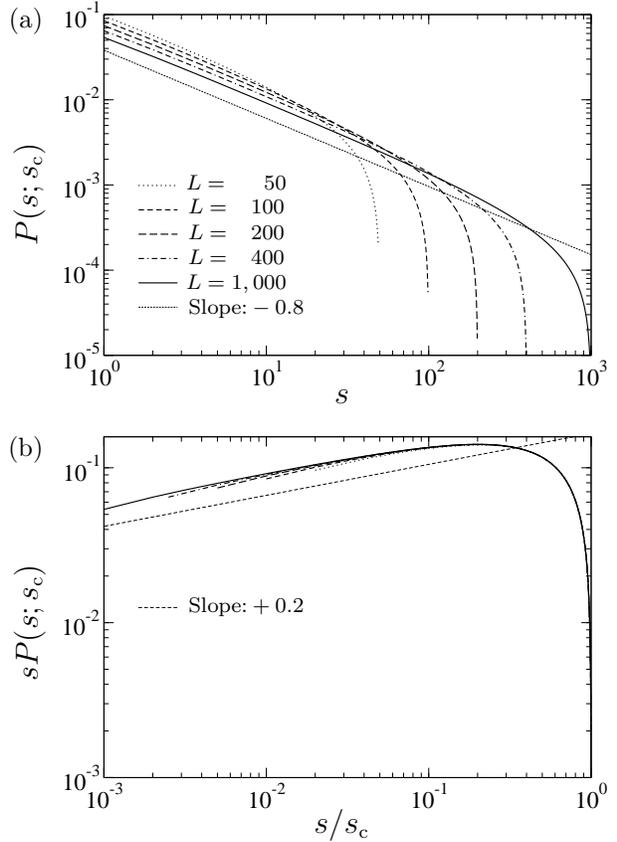}
\caption{
(a) The PDF $P(s;\upCut)$ vs. $s$
with an apparent exponent $\tautilde = 0.8$ (indicated by a straight
line with slope $-0.8$).
Since $\tautilde<1$,
the prefactor $\atilde$ decreases 
with increasing cutoff, shifting the graphs downward the ordinate, 
see Eq.~(\ref{eq:normalisation_condition}b).
(b) Data collapse. The transformed PDF
$s P(s;\upCut)$ vs. 
the rescaled observable $s/\upCut$. For $s,\upCut \gg \loCut$ the curves collapse onto the graph for the scaling
function $\mathcal{G}(x) = x^{1-\tautilde}\GCtilde(x)$
where the cutoff function 
$\GCtilde(x)=(1-x)^{\tautilde}\theta(1-x)$, see \Eref{measured:Ps}. For
small arguments the scaling function behaves like a power law $x^{0.2}$,
as indicated by a straight line with slope $+0.2$.
Example taken from Ref. \cite{FaridChristensen:2006}.
} \flabel{Fig1}
\end{figure}

In this particular example \eref{measured:Ps}, in the limit of large $\upCut$ the scaling
form of the PDF is expected to account for all $s$, so that in this
limit
\begin{eqnarray}
1&=&\int_{\loCut=1}^\infty ds P(s;\upCut) \nonumber \\
&=&\elabel{detatilde}
\left\{
\begin{array}{lr}
\atilde \frac{1}{\tautilde-1} &
\text{for $\tautilde>1$,\qquad (a)}\\
\atilde \upCut^{1-\tautilde} \Gamma(1-\tautilde) \Gamma(1+\tautilde) &
\text{for $\tautilde<1$.\qquad (b)}\\
\end{array}
\right.
\elabel{normalisation_condition}
\end{eqnarray}
For simplicity, we ignore the case
$\tautilde=1$ which contains logarithmic corrections.
\Ereffirst{detatilde} determines $\atilde$, which for $\tautilde<1$ cannot be a
constant but has to vanish like ${\upCut}^{\tautilde-1}$ in the limit of large
$\upCut$. Substituting $\atilde$ into \Eref{measured:Ps} and rearranging, the PDF reads
\begin{eqnarray}
\elabel{Proper_FSS}
&& P(s;\upCut)\\
&=&\left\{ \nonumber
\begin{array}{lr}
(\tautilde\!-\!1) s^{-\tautilde} \GCtilde\left({s}/{\upCut}\right)
&\text{\quad for $\tautilde\!>\!1$,} \\
\!\!\frac{1}{\Gamma(1-\tautilde)\Gamma(1+\tautilde)} s^{-1}
\left(s/\upCut\right)^{1-\tautilde}{\GCtilde}\left({s}/{\upCut}\right)
&\text{\quad for $\tautilde\!<\!1$.} 
\end{array}
\right.
\end{eqnarray}
By comparing \Eref{Proper_FSS} with the FSS ansatz \eref{FSS}, we see that
when the apparent exponent $\tautilde > 1$, the metric factor $a = (\tautilde -1)$, 
the scaling
exponent $\tau = \tautilde$, and the scaling function $\GC(x) = \GCtilde(x)$.
However, if the measured apparent exponent $\tautilde <1$, the
metric factor $a = (\Gamma(1-\tautilde)\Gamma(1+\tautilde))^{-1}$, the scaling exponent
$\tau = 1$, and the scaling function
$\GC(x) = x^{1-\tautilde}\GCtilde(x)$.
For the FSS ansatz \eref{FSS}, it is a necessity that the metric factor is
asymptotically independent of $\upCut$. Dropping this constraint would render
the \emph{definition} of the scaling exponent $\tau$
meaningless: One may rewrite $a s^{-\tau} \GC(s/\upCut) = 
a \upCut^{\alpha} s^{-(\tau + \alpha)} (s/\upCut)^{\alpha} \GC(s/\upCut)$
for any $\alpha$. The factor $(s/\upCut)^{\alpha}$ could be absorbed into the
scaling function. Were it allowed to absorb the power $\upCut^{\alpha}$
in the metric factor, it would be impossible to differentiate between $\tau$ and
$\tau+\alpha$.

\section{Derivation}
Now we present a more general and rigorous derivation of the result
illustrated above.

\subsection{Definitions and notation}
The FSS ansatz \eref{FSS} can be recast more formally as follows:
For all $\epsilon>0$ there exists a constant lower cutoff $\loCut$ and an
$\mathcal{S}$ so that for fixed $n\ge0$, the relative error of the 
$n$th moment from the approximate PDF of the
observable $s\in[0,\infty[$
\begin{subnumcases}{\elabel{FSS_2case}
P_{\text{\tiny app}}(s;\upCut) =}
a s^{-\tau} \GC(s/\upCut) & for $s \geq \loCut$, \elabel{FSS_2universal} \\
f(s,\upCut)               & otherwise
\end{subnumcases}
compared to the exact $n$th moment of the PDF approximated by $P_{\text{\tiny app}}(s;\upCut)$
is less than or equal to $\epsilon$ for all
$\upCut>\mathcal{S}$. Here, we have
introduced $f(s,\upCut)$, the non-universal part of the PDF. 
Both functions, $f(s,\upCut)$ and $\GC(x)$, are non-negative,
and the latter is assumed to be non-zero at least in a finite range of $x$. We
assume $\loCut>0$, but the case $\loCut=0$ is briefly
discussed as well.

The exact $n$th moment is given by
\begin{equation}
\ave{s^n} = f_n(\loCut,\upCut) + a \upCut^{1+n-\tau} g_n(\loCut/\upCut) + \OC(\epsilon)
 \elabel{mom_scaling}
\end{equation}
where
\begin{subequations}
\elabel{def_fn_gn}
\begin{eqnarray}
f_n(\loCut,\upCut) & = & \int_{0}^{\loCut} \dint{s} s^n f(s,\upCut), \\
g_n\left(\loCut/\upCut\right)&=&\int_{\loCut/\upCut}^\infty \dint{x} x^{n-\tau} \GC(x).
\end{eqnarray}
\end{subequations}
Both $f_n$ and $g_n$ are non-negative and $g_n$ is
a monotonically non-increasing function of $\loCut/\upCut$.
It is worth noting that the range over which the scaling
region is probed widens as $\loCut/\upCut$ approaches $0$.

\subsection{Scaling exponent $\tau \geq 1$}
Since $g_0$ does not decrease with $\upCut$, the normalisation condition
\begin{equation}
1=\ave{s^0}=f_0(\loCut,\upCut)+a\upCut^{1-\tau} g_0(\loCut/\upCut)+\OC(\epsilon)
\elabel{normalisation}
\end{equation}
implies $\tau \ge 1$, otherwise $a \upCut^{1-\tau} g_0(\loCut/\upCut)$
would diverge as $\upCut \to \infty$. \result{Therefore, the scaling exponent $\tau \ge 1$.}

\subsection{If $\GC$ is continuous, the limit ${\GC}_0=\lim_{x \to 0}
\GC(x)$ exists and is finite}
The behaviour of the scaling function $\GC$ in the limit of small arguments $s/\upCut$
enters into the scaling of the moments in the asymptotic limit of large upper
cutoff (the thermodynamic limit). 
We will assume that $\GC(x)$
is continuous within a finite range $]0,q]$ with $q > 0$ and that the
integrals in \Eref{def_fn_gn}
exist. Without these (reasonable)
assumptions, the discussion of the behaviour of the integral $g_n$
becomes too complicated. 
The continuity implies further that $\GC(x)$ cannot
diverge with small arguments:
If it did, $g_0(\loCut/\upCut)$ 
diverges faster than $\upCut^{\tau-1}$ for
$\tau>1$ and at least
logarithmically fast for $\tau=1$, contradicting the normalisation condition \Eref{normalisation}. 
\result{Therefore, if $\GC(x)$ is continuous, it cannot be divergent in small arguments.
Together with the continuity, the limit $\GC_0 \equiv \lim_{x \to 0}
\GC(x)$ therefore exists and is finite.}

\subsection{If $\tau = 1$ then $\GC_0 = 0$}
\slabel{tau1_g00}
We consider the case $\GC_0>0$: If $\GC(x)$ is non-zero and bounded from above
for small $x \geq 0$, the integral
$g_n(\loCut/\upCut)$ is logarithmically divergent in $\upCut$ when
$1+n-\tau=0$, has leading order $\upCut^{-(1+n-\tau)}$ when $1+n-\tau<0$
and converges to a non-zero value with increasing $\upCut$ when $1+n-\tau > 0$. The
normalisation condition \Eref{normalisation} implies $a \upCut^{1-\tau} g_0(\loCut/\upCut)\le1$,
so that $\GC_0>0$ violates the normalisation condition if
$\tau \leq1$. For $\tau > 1$, the divergence $s_c^{-(1-\tau)}$ of $g_0$ is cancelled by the
prefactor $s_c^{1-\tau}$. \result{Therefore, if $\GC_0>0$ then $\tau>1$.}
Negating this result yields: \result{If $\tau=1$ then $\GC_0=0$.} 

\subsection{If the lower cutoff $\loCut=0$ then $\tau=1$ and ${\GC}_0=0$}
The result above can be amended by the observation that if the lower cutoff $\loCut=0$ then all
non-universal $f_n$ vanish and the normalisation condition \Eref{normalisation} reduces to $1=a
\upCut^{1-\tau} g_0(0)+\OC(\epsilon)$, which can hold only if $\tau=1$. Moreover,
$g_0(0)=\int_0^\infty \dint{x} x^{-1} \GC(x)$ can converge for continuous $\GC(x)$ 
only if $\GC_0=0$, consistent with our findings in \Sref{tau1_g00}
\result{Therefore, if the lower cutoff vanishes, $\loCut = 0$, then $\tau=1$ and $\GC_0=0$.} Generally,
the converse does not hold, that is, $\GC_0=0$ does not imply
$\loCut=0$. However, $\GC_0>0$ or $\tau>1$ implies $\loCut>0$.

\subsection{If ${\GC}_0=0$ then $\tau =1$}
We consider the case $\GC_0=0$:
Here, we focus
on $g_0$. If $\tau>1$, then $g_0\left(\loCut/\upCut\right)$ diverges slower than
$\left(\loCut/\upCut\right)^{1-\tau}$ in decreasing $\loCut/\upCut$. 
This can be proven by assuming
the opposite, implying that there exists a 
$c>0$ so that $\int_{\loCut/\upCut}^\infty \dint{x}
x^{-\tau} \GC(x)\ge c\left(\loCut/\upCut\right)^{1-\tau}$ for all $\loCut/\upCut$
below a threshold
determined by $c$. The
right hand side of this relation can be written as
$(\tau-1) \int_{\loCut/\upCut}^\infty \dint{x} x^{-\tau}$.
Independent of its prefactor, for
sufficiently small $x$ the integrand
represents an
upper bound, given that $\GC(x)$ is less than any $\delta$ for
sufficiently small $x$. It is easily shown that the above relation
therefore eventually breaks down, implying that the product
$a \upCut^{1-\tau} g_0(\loCut/\upCut)$, the fraction of events within the
universal scaling part of the PDF $P(s;\upCut)$, 
vanishes in increasing $\upCut$ if
$\tau>1$. No such argument exists if $\tau=1$. 
Mathematically, a vanishing fraction of such ``universal events'' constraints
neither $\GC(x)$ nor $\tau$. However, it makes sense to demand some weight
in the scaling part of the PDF on
physical grounds. \result{Therefore, if $\GC_0$ vanishes and the fraction of events
in the scaling part of the PDF is finite in the thermodynamic limit, then
the scaling exponent $\tau = 1$. If $\GC_0$ vanishes and the scaling exponent $\tau > 1$, then
the fraction of events 
in the scaling part of the PDF vanishes in the thermodynamic limit.}
Hence, we are faced with the almost paradoxical situation that all moments
with $n > \tau -1$ are determined by the scaling part of the PDF, the measure
of which, however, vanishes in thermodynamic limit.
In this case, the normalisation condition \Eref{normalisation}
entails that $f_0(\loCut,\upCut)$ converges to $1$ in the limit of large $\upCut$, that is,
the lower cutoff $\loCut$ cannot vanish.
Finally, we note that, if $\GC_0=0$, then $g_n\left(\loCut/\upCut\right)$ cannot
diverge faster for $\loCut/\upCut \to 0$ than in the case $\GC_0>0$ since
$\GC(x)$ does not diverge for small arguments. \result{Therefore, 
if $\GC_0=0$ then $g_n\left(\loCut/\upCut\right)$ converges for all $n>\tau-1$.}

\subsection{If ${\GC}_0 > 0$ then $\tau = \tautilde$ \label{SubSec}}
The apparent exponent $\tautilde$ is formally defined as
\begin{equation}
\lim_{s\to\infty} \lim_{\upCut\to\infty} \frac{P(\lambda s;
\upCut)}{P(s; \upCut)} \lambda^{\tautilde} =1 
\quad \text{for all $\lambda>0$},
\elabel{good_def_tautilde}
\end{equation}
if the limits and such an exponent exist, which is, for
example, not the case if $\GC(x)$ converges to $0$ faster than any power
law.
The limit $s\to\infty$ guarantees that the limit $\upCut\to\infty$ is
taken in the scaling region, so that $P(s;\upCut)$ can be replaced by
the scaling part \Eref{FSS_2universal}:
\begin{equation}
\lim_{s\to\infty} \lim_{\upCut\to\infty} \frac{\GC(\lambda s/\upCut)}{\GC(s/\upCut)}
\lambda^{\tautilde-\tau} =1 \quad \text{for all $\lambda>0$}.
\elabel{GC_lim}
\end{equation}
If $\GC_0>0$, the fraction in
\Eref{GC_lim} converges to $1$ and for the left hand side to be equal to unity for all 
$\lambda$, the exponent $\tautilde-\tau$ vanishes. \result{Therefore, if
$\GC_0>0$
then $\tau = \tautilde$ or, equivalently,
$\tau \neq \tautilde \Rightarrow \GC_0=0$.}

\subsection{Generally $\tau \geq \tautilde$, 
            if $\tautilde < 1$ then $\tau=1$, 
	    and if $\tautilde \ge 1$ then $\tau=\tautilde$}
\Ereffirst{GC_lim} implies that $\GC(\lambda s/\upCut)$ scales like
$\lambda^{\tau-\tautilde}$ and therefore diverges in small arguments if
$\tau < \tautilde$. However, as noted previously, $\GC$ cannot diverge for small arguments.
\result{Therefore, we have the general result $\tau \ge \tautilde$.}

Moreover, the cutoff function $\GCtilde(x)=x^{\tautilde-\tau}\GC(x)$
can then be expected to converge to a (non-vanishing) constant as $x \to 0$, and imposing that,
one can define uniquely the apparent exponent $\tautilde$ through
\begin{equation}
P(s;\upCut) = \atilde(\upCut) s^{-\tautilde}  \GCtilde\left(s/\upCut\right) \quad\text{for $s \geq \loCut$}.
\elabel{Def:measured:Ps}
\end{equation}

Using the above two
arguments, the scaling exponent $\tau$ and the apparent exponent $\tautilde$
can be related as follows: If
$\tautilde<1$ then $\tau \ne \tautilde$ since $\tau\ge1$, implying
$\GC_0=0$. As shown above,
$\GC_0=0$ implies $\tau=1$ assuming that the scaling part of the PDF
is non-empty. \result{Therefore, if $\tautilde<1$ then $\tau=1$.} 

Now, assume to the contrary that $\tau \ne \tautilde \ge1$. Hence
$\GC_0=0$ and again, assuming a non-empty scaling part of the PDF, $\tau=1$. If
$\tautilde=1$ that clashes with $\tau\ne\tautilde$, if $\tautilde>1$
that clashes with $\tau \ge \tautilde$. 
\result{Therefore, if $\tautilde \ge 1$ then $\tau = \tautilde$,
unless the total measure of the scaling part of the PDF vanishes
asymptotically.}

\section{Discussion}
In this section, we first discuss some implications of the above
derivations and then present some examples from the literature, where a
more thorough scaling analysis leads to revised results.

In many models of equilibrium
\cite{PfeutyToulouse:1977} as well as non-equi\-lib\-ri\-um
\cite{MarroDickman:1999} statistical
mechanics, within the FSS regime, the second moment $\langle s^2 \rangle$ of the
total order
parameter\footnote{We distinguish between the total order parameter $s$
and the order parameter density $s/L^d$. The analysis is identical for the order parameter
density, however, the upper cutoff then vanishes asymptotically which
would require a revision of previous arguments.}
scales like its first moment squared ${\langle s \rangle}^2$, which is usually linked to
so-called hyper-scaling. This implies that their ratio converges in the
thermodynamic limit, that is,
\begin{equation}
\lim_{\upCut \to \infty} \frac{\ave{s^2}}{\ave{s}^2} = r,
\elabel{mom_rat}
\end{equation}
where $r$ is a finite number.
We now show that such finite $r$ implies $\tau=1$, 
assuming that  \Eref{FSS} holds in these
models, by showing that if $\tau>1$, then the ratio
diverges asymptotically. To invoke the argument ``if $\tau>1$ then
$\GC_0>0$'', we need to impose that a non-vanishing fraction of the
total order parameter distribution is governed by finite-size scaling.
In equilibrium statistical mechanics, that means that the singular part
of the partition sum does not vanish asymptotically.
Moreover, we assume that both the first and the second moment diverge
in the thermodynamic limit. This is important, because $r$ may be finite
even for $\GC_0>0$ simply by convergence of the first two moments if
$\tau>3$, as can be shown using \Eref{mom_scaling} and the divergence of
$g_n$ as discussed in Sec.\ref{SubSec}. Based
on the arguments presented there, if the first moment diverges, then
$\tau\le2$. Using the same arguments, one can now show that for
$2>\tau>1$ the ratio $r$ diverges like $\upCut^{\tau-1}$ and for
$\tau=2$ it diverges like $\upCut^{\tau-1}$ with logarithmic
corrections. Thus asymptotically finite $r$ is incompatible with
$\tau>1$, if the first moment diverges and the fraction of the total order
parameter distribution that is governed by the scaling part is finite.
The only alternative is $\tau=1$, in which case $\GC_0=0$ and therefore
all $g_n$ are finite for all $n$ and $r=a^{-1} g_2/g_1^2$ from
\Eref{mom_scaling}. The case $\tau=1$ therefore is anything but special,
describing most of standard critical phenomena. It is worth noting that
$\tau=1$ implies a dimensionless metric factor $a$.

We illustrate the observation above by considering the finite-size scaling of
the cluster number density 
\footnote{In the
following, we consider finite systems at occupation probability $p$
tuned to the
critical value $p_c$, so that we do not have to exclude from the
cluster number density the largest,
infinite or percolating cluster(s).}
in percolation \cite{StaufferAharony:1994} (i.e., the number of clusters
of size $s$ per site in a system of linear size $L$)
\begin{equation}
n(s;\upCut^{\perc})= a_{\perc} s^{-\tau^{\perc}} \GC^{\perc}\left(s/\upCut^{\perc}\right),
\elabel{perc_FSS}
\end{equation}
where $\upCut^{\perc} = b^{\perc} L^D$, $D=1/(\nu\sigma)$ with $\nu$ and $\sigma$ being the critical
exponents describing the divergence of the correlation length and the characteristic
cluster size, respectively, and $b^{\perc}$ being a metric factor.
The exponent $\tau^{\perc}$ is
different from $1$, known to be exactly $187/91$ in two-dimensional
percolation. However, $n(s;\upCut^{\perc})$ is not the distribution of the total order
parameter, which is the number of occupied sites in the largest
cluster. Multiplying $n(s;\upCut^{\perc})$ by $L^d$, one obtains the average number of
clusters of size $s$ in a system with $L^d$ sites. The probability of the
largest cluster being of a particular size must be contained in this
distribution: If $n(s;\upCut^{\perc})$ was the site-normalised density of the largest
cluster only, then $L^d n(s;\upCut^{\perc})$ was the probability that the largest
cluster has size $s$. If $\tilde{n}(s;\upCut^{\perc})$ is the site-normalised histogram of all clusters
excluding the largest, one could therefore write $L^d n(s;\upCut^{\perc}) = L^d
\tilde{n}(s;\upCut^{\perc}) + P(s;\upCut^{\perc})$ with
$P(s;\upCut^{\perc})$ being the distribution
of the total order parameter. One might speculate that
$P(s;\upCut^{\perc})$ is
responsible for the characteristic bump in the histogram
$n(s;\upCut^{\perc})$.
\emph{Assuming} that $P^{\perc}(s;\upCut^{\perc})$ follows the same scaling as $L^d
n(s;\upCut^{\perc})$ itself, one has
\begin{equation}
P^{\perc}(s;\upCut^{\perc}) = a'_{\perc} L^d s^{-\tau^{\perc}}
\GCtilde^{\perc}(s/\upCut^{\perc}),
\end{equation}
with $\upCut^{\perc}= b^{\perc} L^D$. Absorbing $L^d$ into a redefinition of
the scaling function $\GC^{\perc}(x)=x^{-d/D}
\GCtilde^{\perc}(x)$ 
the distribution of the order
parameter scales like
\begin{equation}
P^{\perc}(s;\upCut^{\perc}) = a'_{\perc} s^{\frac{d}{D}-\tau^{\perc}}
\GC^{\perc}(s/\upCut^{\perc}),
\end{equation}
where the exponent of $s$ turns out to be $d/D-\tau^{\perc}=-1$
by standard scaling laws \cite{StaufferAharony:1994}, indeed as expected
from the previous analysis.

\subsection{Examples}
Finally, we want to present some examples from the literature, where a
misunderstanding of the theory of scaling has lead to some confusion or
even erroneous conclusions.
In \cite{RamstadHansen:2006} the cluster number density in
steady state two-phase flow in porous media is investigated. As
discussed above, in percolation the
cluster number density $n(s)$ is the
site normalised density of clusters of size $s$, so that $s n(s)$ is the
probability that a randomly chosen site belongs to a cluster of size $s$.
The histogram is expected to follow scaling, $n(s) = a^{\perc}
s^{\tau^{\perc}} \GC^{\perc}(s/\upCut^{\perc})$, see \Eref{perc_FSS}. 
The sum $p=\sum_s s n(s)$ is the fraction of sites belonging to any
cluster, \ie the fraction of occupied sites.
This sum is bounded and non-zero in non-trivial cases even if the
largest clusters are excluded from $n(s)$, so that the
results for $P(s;s_c)$ derived in this paper, 
translate into corresponding results for $s n(s)$.
Performing the sum suggests $p\propto
(\upCut^{\perc})^{2-\tau^{\perc}}$, which means that 
$p$ vanishes asymptotically for
$\tau^{\perc}>2$. This, however, is a misunderstanding
of the theory of scaling, because it ignores the existence of a finite
lower cutoff $\loCut^{\perc}$. In fact, $\tau^{\perc}\ge2$,
corresponding to $\tau\ge1$ as derived earlier, is a
necessary condition for $p$ to be finite. The numerical finding of
$\tau^{\perc} \approx 1.92$ in \cite{RamstadHansen:2006} for a cluster number density
therefore is theoretically impossible -- in fact it is based on a
straight line fit in a double logarithmic plot and therefore it is
$\tautilde$ and not $\tau$ that is estimated.

The term self-organised criticality refers to the tendency of
slowly driven non-equilibrium
systems with many degrees of freedom to spontaneously reach a
critical state where
the PDF of relaxational event sizes obey FSS. 
There are only a few controlled laboratory
experiments on scale-invariance in slowly-driven non-equilibrium systems.
In one such experiment, the avalanche-size PDF was measured in slowly driven piles of rice \cite{Ricepile:1996}.
For piles with elongated grains, the avalanche-size PDF obeys FSS. However, it was claimed that
a system with more spherical grains is non-critical, since the PDF of the avalanche size $s$ in a
system of size $L$ is consistent with $P(s;L) \propto L^{-1} \GCtilde(s/L)$ where the
function $\GCtilde(x) \propto \exp\left[-(x/x^{\star})^{\gamma}\right]$ with
$\gamma \approx 0.43$ and $x^{\star} \approx 0.45$ was identified as a scaling function rather than a
cutoff function. This is a misunderstanding of the theory of scaling.
The PDF can be recast into
the FSS form $P(s;L) \propto s^{-1} \GC(s/L)$ where the scaling function
$\GC(x) = x \GCtilde(x)$ increases linearly with $x$ for small
arguments, decaying (stretched) exponentially fast for large arguments. Hence, also for the piles with
more spherical grains does the avalanche-size PDF obey FSS but with a scaling
exponent $\tau = 1$ and 
$\GC_0 = 0$.
Therefore, the occurrence of
self-organised criticality does not depend on the details of the system
as wrongly claimed in \cite{Ricepile:1996}.

Complex networks have attracted intense interest from the statistical
physics community in particular because they often exhibit scale invariance.
It has been suggested that so-called
du\-pli\-ca\-tion-di\-ver\-gence-mu\-ta\-tion models,
where nodes correspond to proteins and links correspond to pairwise
interactions between proteins,
are good candidates to describe the evolution and large-scale topological
features of real protein-protein interaction networks.
A recent study \cite{Raval:2003} derives asymptotic properties for such models and it is
reported that the degree PDF for a node to have degree $k$ at time $t$
is $P(k;t) \propto t^{-1} (k/t)^{k_{\text{\tiny min}}-1} \GCtilde(k/t)$
where $k_{\text{\tiny min}}$
is the lower non-zero degree in the initial configuration, $\GCtilde$ a cutoff function, and
it is claimed that the scaling exponent is $k_{\text{\tiny min}}-1$ \cite{Raval:2003}. However,
this is a misunderstanding of the theory of scaling. A careful analysis reveals that
the PDF can be recast into the FSS form $P(k;t) = a k^{-1} \GC(k/t)$ where the
metric factor $a = k_{\text{\tiny min}}(1+k_{\text{\tiny min}})$ and
the scaling function
$\GC(x) = x^{k_{\text{\tiny min}}} (1-x) \theta(1-x)$ increases like a power law with exponent
$k_{\text{\tiny min}}$ with $x$ for small arguments, decaying fast for large arguments.
Hence, the FSS scaling exponent $\tau = 1$, independent of
$k_{\text{\tiny min}}$ and
$\GC_0 = 0$. 

Earthquake statistics has been a prominent subject of statistical analysis
by means of scaling arguments (\eg \cite{BakETAL:2002}). In
\cite{DavidsenPaczuski:2005}, the
distribution $P_{m,L}(\Delta r)$ of distances of successive earthquakes with a
magnitude greater than $m$ within a cell of linear $L$ is considered.
The analysis shows that $P_{m,L}(\Delta r) = L^{-1} f(\Delta r/L)$,
where $f(x) \propto x^{-0.6}$. As that implies $P_{m,L}(\Delta r)
\propto \Delta r^{-0.6}$, 
the distribution seems to be asymptotically non-normalisable.
This conclusion, however, is a misunderstanding of the theory of scaling
and a confusion of $\tautilde$ and $\tau$: 
The PDF $P_{m,L}(\Delta r)$ was non-normalisable only if $\tau<1$, but
in fact $\tau=1$, which can be seen by
rewriting the scaling of the PDF as $P_{m,L}(\Delta r)=\Delta r^{-1} \GC(\Delta
r/L)$ with $\GC(x)=x f(x)$. The scaling function $\GC(x)$ converges to $0$ for
small arguments like $x^{1-0.6}$, as expected from the discussion above.
In the introduction we showed that $\GC(x)\propto x^\alpha$ implies 
$\tautilde=\tau-\alpha$, so that in fact $\tautilde=0.6$ in the present
case.

In summary, given the probability density function $P(s;\upCut)$ obeys the FSS ansatz \eref{FSS},
if the apparent exponent $\tautilde$ of $P(s;\upCut)$ in \Eref{power-law} is less than unity, then
the scaling exponent $\tau$ is unity and the scaling function vanishes for
small arguments, $\lim_{x \to 0} \GC(x) = 0$.
If $\tautilde \ge 1$, then the scaling exponent equals the
apparent exponent, $\tau = \tautilde$. If $\tau > 1$, then the scaling function approaches a non-zero value for
small arguments, $\lim_{x \to 0} \GC(x) > 0$.

\begin{acknowledgement}
GP thanks Mervyn Freeman, Nicholas Watkins and Max Werner for useful
discussions.
\end{acknowledgement}

\bibliography{ScalingFunction}
\end{document}